\documentclass{article}

\usepackage{amssymb}
\usepackage{amsmath}
\usepackage{mathrsfs} 
\usepackage[all]{xy}
\usepackage[dvips]{color}
\usepackage[dvips]{graphicx}

\usepackage{hyperref}

\definecolor{dyellow}{rgb}{1.,0.8,.0}
\definecolor{myblue}{rgb}{.1,.1,.7}
\definecolor{dcyan}{rgb}{.0,.6,.6}
\definecolor{dmagenta}{rgb}{0.6,0.0,0.6}
\definecolor{brown}{rgb}{0.6,0.2,0.}
\definecolor{darkblue}{rgb}{.0,.0,0.5}
\definecolor{darkred}{rgb}{0.75,0.0,0.0}
\definecolor{orange}{rgb}{1.,.6,.0}
\definecolor{dorange}{rgb}{0.8,.4,.0}
\definecolor{darkgreen}{rgb}{0.0,0.6,0.0}
\definecolor{purple}{rgb}{.4,.0,.4}
\definecolor{grey}{rgb}{0.7,0.7,0.7}

\newcommand{\hide}[1]{}
\newcommand{\delete}[1]{}
\newcommand{\distribute}[3]{}	

\newcommand{\vect}[1]{\ensuremath{\boldsymbol{#1}}}
\newcommand{\tensor}[1]{\mathbf{#1}}

\newcommand{\od}{\mathrm{d}}

\newcommand{\Lied}[1]{L_{#1}}

\title{U(1) Gauge Potentials on de Sitter Spacetime}
\author{Bin Zhou\thanks{Email: \texttt{zhoub@bnu.edu.cn}}, $^{1}$
Shi-Bei Kong\thanks{Email: \texttt{kongshibei@mail.bnu.edu.cn}}, $^{1}$
and Peng Zhao\thanks{Email: \texttt{200911131058@mail.bnu.edu.cn}}, $^{2}$ 
\\
$^{1}$ Department of Physics, Beijing Normal University, Beijing 100875, China
\\
$^{2}$ Department of Mathematics, Beijing Normal University, Beijing 100875, China}

\begin{document}

\maketitle

\begin{abstract}

The smooth 1-form Verma module of $\mathfrak{so}(1,4)$ is acquired, which can be regarded as the U(1) gauge potential on de Sitter spacetime. 
It is shown that electromagnetic fields could not be source free on de Sitter background.

\end{abstract}


\section{Introduction}

Classical electrodynamics, which was developed mostly by James Clerk Maxwell,
is one of the greatest theories ever since Newton's mechanics.
In classical electrodynamics, there are some key points concerned in this paper:
\begin{itemize}
  \item[(1)] The existence of electromagnetic waves were at first predicted theoretically,
  then and confirmed experimentally by Heinrich Hertz later;
  \item[(2)] Electromagnetic fields carry energy and momentum as well as
  ordinary matters. The energy-momentum tensor of ordinary matters is not conserved, and
  the energy-momentum tensor of EM fields is not conserved,neither.However, the energy-momentum tensor of ordinary matters
 and electromagnetic fields combined is a conservative quantity.
  \item[(3)] There are nontrivial solutions of source-free Maxwell equations,
 which means that electromagnetic fields could exist independent of charges and
  currents.
  \item[(4)]In this way Electromagnetic fields are recognized as
  an existing special form of matter not rather than just an imagination or mathematical concept.
  \item[(5)] In vacuum, electromagnetic waves propagate at a universal speed
  $c$having nothing to do irrelevant with time, space point, propagation direction,
  as well as the state of the emitting source. The speed $c$ is that equivalent to the speed of light
  in vacuum.  Thus James C. Maxwell concluded that light is nothing but
  a kind of electromagnetic waves, hence and is also dominated by the Maxwell equations.
  \item[(6)] It also seems that the speed of light, in vacuum, is independent of different choices of 
  (inertial) reference frames.
\end{itemize}

The validity of point~(6) was an open question before 1905.
This question, together with the problem of the covariance of electrodynamics,
finally resulted in the discovery of the special theory of relativity (SR),
and then the general theory of relativity (GR).  Especially, point~(6)
is recognized by Albert Einstein as a fundamental postulate of SR,
the principle of the invariance of light speed.

In SR, electrodynamics is well adapted to the relativistic concept of
spacetime, with all the above points perfectly preserved.

In GR, most of dynamical equations (except Einstein's field equation and so on)
are modified from those in SR.  For a long time it is thought that the consequence
of relativistic dynamics remains still valid in GR, at least remained
in a modified form.  For examples, points~(1) to (4) in the above are thought
to be still correct in GR, while points~(5) and (6) are thought to be valid
in a modified form.

Speed of light in vacuum. In general relativity, giving solutions of sourceless electromagnetic waves.
Physical laws.

In this paper, we show, by virtue of an example, that it is not the case:
basic knowledge obtained from SR is not necessarily valid in GR.
To be specific, we shall show that there should could be no source-free
electromagnetic fields in the de~Sitter background.  In SR, the distribution
of (electrical/magnetic) charges and currents cannot determine
the electromagnetic fields.  To determine the latter, initial condition and
boundary conditions are needed.  In the de~Sitter background, however,
initial condition and boundary conditions are not necessary; the distribution of
(electric/magnetic) charges and currents is enough to determine
the electromagnetic field.

The absence of source-free electromagnetic fields in de~Sitter background
brings out a serious problem: the experimental foundation of the curved metric.
In GR, the numerical value of light speed might somehow be meaningless.
However, the 4-``velocity'' of a photon in vacuum makes sense, which is specified
as a lightlike vector with respect to the metric, and vice versa.
By means of lightlike vectors of various directions, the spacetime metric can be
determined up to a conformal factor.  This is the experimental foundation of
the curved metric in GR.
If there is no source-free electromagnetic fields, there is no way to obtain
lightlike 4-vectors. 
In this case the curved metric will lose its experimental foundation.

In this paper we are not going to discuss such a problem.  We mainly focus on
whether there is source-free electromagnetic fields in the de Sitter background.
The fundamental technique to solve this problem is the application of theory of
Lie groups and Lie algebras.

This paper is organized as the following.  In section~\ref{sect:prelim}
we outline the ideas and present some preliminary formulae.
In section\ref{sect:highest-vector} vector fields,
as the highest weight vector in the representation of the Lie algebra
$\mathfrak{so}(1, 4)$, are obtained.  Following the standard method,
Verma modules of $\mathfrak{so}(1,4)$, consisting of vector fields, or 1-forms,
or 3-forms, respectively, are obtained.  These Verma modules are irreducible
$\mathfrak{so}(1, 4)$-modules.  They are described
in section\ref{sect:Verma-modules}.
In section\ref{sect:properties} some important properties of 1-forms
in the above modules are discussed.  These are applicable in constructing
smooth solutions of the Maxwell equations on $dS^{4}$, or smooth solutions
of the Proc equation.  These solutions are
described in section\ref{sect:smooth-solutions} and section\ref{sect:mass-Proc}, respectively.
In section\ref{sect:CD} we summarize the whole paper, and discussed some
related problems.

This work is a natural extension of the previous work by Bin Zhou and Zhenhua Zhou, where the necessary mathematical preparations, such as
the expressions of the root vectors of the Lie algebra so(1,4) have been made.
In that paper, we have got the simplest representations of the algebra, which is the Verma module of smooth functions, and then the solutions of
Klein-Gordon equation. It's perfectly natural for us to find the Verma module of higher rank tensors,especially forms,which are anti-symmetric, and use them
to find the solutions of other field equations,such as Maxwell's equations,Proca equation even Dirac equation.
Apparently,we are not the first to investigate this problem, but our method is novel as far as we know.
Most of the previous work could not get an exact solution and used some special functions such as hyper-geometric function which is very complicated.
What's more, they could not specify the relationship of their solution with the symmetry of the de Sitter spacetime itself.

\section{Some Preliminary Formulae}\label{sect:prelim}

The coordinate system $(\chi, \zeta, \theta, \varphi)$ on $dS^{4}$ is defined by
\begin{alignat}{1}
  \xi^{0} &= l \sinh\chi
  \, ,
  \label{eq:coord-sys:0}
  \\
  \xi^{1} &= l \cosh\chi \cos\zeta \cos\theta
  \, , \\
  \xi^{2} &= l \cosh\chi \cos\zeta \sin\theta
  \, , \\
  \xi^{3} &= l \cosh\chi \sin\zeta \cos\varphi
  \, , \\
  \xi^{4} &= l \cosh\chi \sin\zeta \sin\varphi
  \, .
  \label{eq:coord-sys:4}
\end{alignat}

\section{Highest Weight Vector Fields}\label{sect:highest-vector}

We first start to find a vector field
\begin{equation}
  \tensor{v}_{\lambda} = v^{0} \frac{\partial}{\partial \chi}
  + v^{1} \frac{\partial}{\partial \zeta}
  + v^{2} \frac{\partial}{\partial \theta}
  + v^{3} \frac{\partial}{\partial \varphi}
\end{equation}
on the de Sitter spacetime, acting as a highest weight vector
in the representation of $\mathfrak{so}(1, 4)$ with the weight
$\lambda = N_{1} \lambda_{1} + N_{2} \lambda_{2}$.
The action of $\mathfrak{so}(1, 4)$ on vector fields is realized
by Lie derivatives.  Then $\tensor{v}_{\lambda}$ satisfies
\begin{alignat}{2}
  \Lied{\tensor{h}_{\alpha_{1}}} \tensor{v}_{\lambda}
  &= [\tensor{h}_{\alpha_{1}}, \tensor{v}_{\lambda}]
  = N_{1} \tensor{v}_{\lambda}
  \, , &
  \Lied{\tensor{h}_{\alpha_{2}}} \tensor{v}_{\lambda}
  &= [\tensor{h}_{\alpha_{2}}, \tensor{v}_{\lambda}]
  = N_{2} \tensor{v}_{\lambda}
  \, ,
  \label{eq:v-lambda:h1-h2}
  \\
  \Lied{\tensor{e}_{\alpha_{1}}} \tensor{v}_{\lambda}
  &= [\tensor{e}_{\alpha_{1}}, \tensor{v}_{\lambda}]
  = 0
  \, , & \qquad
  \Lied{\tensor{e}_{\alpha_{1} + 2\alpha_{2}}} \tensor{v}_{\lambda}
  &= [\tensor{e}_{\alpha_{1} + 2 \alpha_{2}}, \tensor{v}_{\lambda}]
  = 0
  \, ,
  \label{eq:v-lambda:e1-e122}
  \\
  \Lied{\tensor{e}_{\alpha_{2}}} \tensor{v}_{\lambda}
  &= [\tensor{e}_{\alpha_{2}}, \tensor{v}_{\lambda}]
  = 0
  \, , &
  \Lied{\tensor{e}_{\alpha_{1} + \alpha_{2}}} \tensor{v}_{\lambda}
  &= [\tensor{e}_{\alpha_{1} + \alpha_{2}}, \tensor{v}_{\lambda}]
  = 0
  \, .
  \label{eq:v-lambda:e2-e12}
\end{alignat}
Eqs.~(\ref{eq:v-lambda:h1-h2}) result in
\[
  i \, \frac{\partial v^{\mu}}{\partial \theta}
  - i \, \frac{\partial v^{\mu}}{\partial \varphi}
  = N_{1} \, v^{\mu}
  \, , \qquad
  2i \, \frac{\partial v^{\mu}}{\partial \varphi} = N_{2} \, v^{\mu}
  \, .
\]
The general solution of the above equations is
\begin{displaymath}
  v^{\mu} = V^{\mu}(\chi, \zeta) \,
    e^{- i (N_{1} + \frac{N_{2}}{2} \theta) - i \frac{N_{2}}{2} \varphi}
  \, ,
\end{displaymath}
where $V^{\mu}(\chi, \zeta)$ are some functions depending only on $\chi$ and
$\zeta$.  Hence
\begin{equation}
  \tensor{v}_{\lambda}
  = e^{- i (N_{1} + \frac{N_{2}}{2}) \theta - i \frac{N_{2}}{2} \varphi} \Big(
      V^{0}(\chi, \zeta) \, \frac{\partial}{\partial \chi}
      + V^{1}(\chi, \zeta) \, \frac{\partial}{\partial \zeta}
      + V^{2}(\chi, \zeta) \, \frac{\partial}{\partial \theta}
      + V^{3}(\chi, \zeta) \, \frac{\partial}{\partial \varphi}
    \Big)
  \, .
\end{equation}
According to eqs.~(\ref{eq:coord-sys:0}) to (\ref{eq:coord-sys:4}),
the spacetime point returns back to its initial position whenever $\varphi$
increases by $2 \pi$.  The above expression implies that $N_{2}$ must be
an even integer.  Soon we shall see that $N_{2}$ could be either zero or 2.

In fact, we can substitute the above expression into
eqs.(\ref{eq:v-lambda:e1-e122}), obtaining eight equations for four unknown
functions $V^{\mu}(\chi, \zeta)$. Among these equations, there are
four PDEs and four linear algebraic equations.  All these equations force
$N_{2}$ to be either $\pm 2$ or 0.

For $N_{2} = 0$, these equations are reduced to
\begin{alignat*}{1}
  & \frac{\partial V^{0}}{\partial \zeta}
  + N_{1} \, V^{0} \tan\zeta
  = 0
  \, ,
  \\
  &
  \frac{\partial V^{2}}{\partial \zeta}
  + (N_{1} - 2) \, V^{2} \tan\zeta
  = 0
  \, ,
\end{alignat*}
together with
\begin{displaymath}
  V^{1} = -i V^{2} \sin \zeta \cos\zeta
  \, , \qquad
  V^{3} = 0
  \, .
\end{displaymath}
The general solution of these equations can be easily obtained, resulting in
the corresponding vector field $\tensor{v}_{\lambda}$ to be
\begin{alignat*}{1}
  \tensor{v}_{N \lambda_{1}}
  \distribute{8}{September 16, 2012}{
  &= e^{-i N \theta} \Big(
      \mathcal{V}^{0} \, (\cos\zeta)^{N} \frac{\partial}{\partial \chi}
      - i \mathcal{V}^{2} \sin\zeta \, (\cos\zeta)^{N - 1}
        \frac{\partial}{\partial \zeta}
      + \mathcal{V}^{2} (\cos\zeta)^{N - 2} \frac{\partial}{\partial \theta}
    \Big)
  \nonumber \\
  }
  &= e^{- i N \theta} \cos^{N} \zeta \,
    \Big(
      \mathcal{V}^{0} \, \frac{\partial}{\partial \chi}
      - i \mathcal{V}^{2} \tan\zeta \, \frac{\partial}{\partial \zeta}
      + \mathcal{V}^{2} \sec^{2} \zeta \, \frac{\partial}{\partial \theta}
    \Big)
  \, ,
\end{alignat*}
where $N_{1}$ has been denoted simply by $N$.
Substitution of this expression into the first equation in
(\ref{eq:v-lambda:e2-e12}) yields
\begin{alignat}{1}
  & \frac{\od \mathcal{V}^{2}}{\od \chi}
  - (N - 2) \, \mathcal{V}^{2} \tanh\chi
  = 0
  \, ,
  \nonumber \\ &
  \mathcal{V}^{0}
  = - i \mathcal{V}^{2} \, \sinh\chi \, \cosh\chi
  \, .
\end{alignat}
The general solution of the above equations is
\[
  \mathcal{V}^{2} = C \, (\cosh \chi)^{N - 2}
  \, , \qquad
  \mathcal{V}^{0} = -i C \, \sinh\chi \, (\cosh\chi)^{N - 1}
  \, ,
\]
where $C$ is the integral constant.  We can fix $C$ as $i N/l$
(where $l$ is the cosmological radius of $dS^{4}$)
so that
\begin{alignat}{1}
  \tensor{v}_{N \lambda_{1}}
  &= \frac{N}{l} \, \phi_{N \lambda_{1}} \,
    \Big(
      \tanh\chi \, \frac{\partial}{\partial \chi}
      + \frac{\tan\zeta}{\cosh^{2} \chi} \, \frac{\partial}{\partial \zeta}
      + i \frac{\sec^{2} \zeta}{\cosh^{2}} \,
        \frac{\partial}{\partial \theta}
    \Big)
  \, ,
\end{alignat}
which satisfies
\begin{equation}
  g_{ab} v_{N \lambda_{1}}^{b}
  = l \, (\od \phi_{N \lambda_{1}})_{a}
  \, .
  \label{gab-vb:N-lambda1}
\end{equation}
Here
\begin{equation}
  \phi_{N \lambda_{1}} = \phi_{\lambda_{1}}^{N}
  = (e^{- i \theta} \cosh\chi \cos\zeta)^{N}
\end{equation}
is the highest weight scalar with the weight $N \lambda_{1}$. For details of
this function, we refer to \cite{Zhou:2011dr}. Since the function
$\phi_{N \lambda_{1}}$ is smooth on $dS^{4}$, so is the 1-form
$\od\phi_{N \lambda_{1}}$.
As a consequence, $\tensor{v}_{N \lambda_{1}}$ is a smooth vector field
on $dS^{4}$.

It can be verified that the second equation in (\ref{eq:v-lambda:e2-e12})
is automatically satisfied.

For the case of $N_{2} \neq 0$, those eight equations forces $V^{0} = 0$.
In order that one of $V^{1}$, $V^{2}$ and $V^{3}$ is nonzero in these equations,
$N_{2}$ must be $\pm 2$.  But according to the representation theory,
$\lambda = N_{1} \lambda_{1} + N_{2} \lambda_{2}$ is a dominant weight,
so that $N_{2} = 2$.\footnote{
When $N_{2} = -2$, one can try to solve the equations.  Then only a zero 1-form
could be obtained.
}
Then those equations, derived from
eqs.~(\ref{eq:v-lambda:e1-e122}), are reduced to
\begin{alignat*}{1}
V^{0} = 0, \quad \frac{\partial V^{1}}{\partial \zeta} + N_{1}, \quad V^{1} \tan\zeta = 0, \quad V^{2} = i V^{1} \tan\zeta, \quad V^{3} = -i V^{1} \cot\zeta.
\end{alignat*}
Having easily obtained the general solutions of these equations, there is
the corresponding
\begin{alignat}{1}
  \tensor{v}_{\lambda}
  &= e^{-i (N + 1) \theta - i \varphi}
    \, \mathcal{V}(\chi) \, (\cos\zeta)^{N}
    \Big(
      \frac{\partial}{\partial \zeta}
      + i \tan\zeta \, \frac{\partial}{\partial \theta}
      - i \cot\zeta \frac{\partial}{\partial \varphi}
    \Big)
  \nonumber \\
  &= 2 \mathcal{V}(\chi) \, (e^{- i \theta} \cos\zeta)^{N}
    \tensor{e}_{\alpha_{1} + 2 \alpha_{2}}
\end{alignat}
with the highest weight $\lambda = N \lambda_{1} + 2 \lambda_{2}$.
In this expression, $\mathcal{V}$ is a function depending on $\chi$ only.
For the expression of $\tensor{e}_{\alpha_{1} + 2\alpha_{2}}$, we refer to \cite{Zhou:2011dr}. 
It can be calulated that
\begin{alignat*}{1}
  \Lied{\tensor{e}_{\alpha_{2}}} \tensor{v}_{\lambda}
  &= 2 e^{- i N \theta - i \varphi}
    \Big( \frac{\od\mathcal{V}}{\od\chi} - N \mathcal{V} \tanh\chi \Big)
    \sin\zeta \, (\cos\zeta)^{N}
    \, \tensor{e}_{\alpha_{1} + 2 \alpha_{2}}
  \, .
\end{alignat*}
Therefore the first equation in (\ref{eq:v-lambda:e2-e12}) requires that
\[
  \frac{\od\mathcal{V}}{\od\chi} - N \mathcal{V} \tanh\chi = 0
  \, ,
\]
which has the general solution $\mathcal{V} = C (\cosh\chi)^{N}$.
With the integral constant $C$ fixed as $1/l$, we can obtain a highest weight
vector field
\begin{alignat}{1}
  \tensor{v}_{\lambda}
  = \frac{2}{l} \, \phi_{N \lambda_{1}} \,
    \tensor{e}_{\alpha_{1} + 2 \alpha_{2}}
  \, , \qquad
  (\lambda = N \lambda_{1} + 2 \lambda_{2})
  \, .
  \label{eq:v-lambda}
\end{alignat}
Now that both $\phi_{N \lambda_{1}}$ and
$\tensor{e}_{\alpha_{1} + 2 \alpha_{2}}$
are smooth on $dS^{4}$ \cite{Zhou:2011dr}, obviously $\tensor{v}_{\lambda}$ is
a smooth vector field on $dS^{4}$.

By virtue of eq.(\ref{eq:v-lambda}) as well as the relations in \cite{Zhou:2011dr}, it is obvious that the second equation
in (\ref{eq:v-lambda:e2-e12}) is automatically satisfied.

\distribute{7}{September 16, 2012}{
There are some useful relations:
\begin{alignat}{4}
  [e^{i \varphi} \tensor{e}_{\alpha_{2}}, \tensor{v}_{*}]
  &= && e^{i \theta} \tensor{e}_{\alpha_{1} + \alpha_{2}}
  \, , &
  [\tensor{e}_{\alpha_{2}}, \tensor{v}_{*}]
  &= && e^{- i \varphi} \tanh\chi \csc\zeta \, \tensor{v}_{*}
  \, ,
  \\
  [e^{i \theta} \tensor{e}_{\alpha_{1} + \alpha_{2}}, \tensor{v}_{*}]
  &= - && e^{i \varphi} \tensor{e}_{\alpha_{2}}
  \, , \qquad &
  [\tensor{e}_{\alpha_{1} + \alpha_{2}}, \tensor{v}_{*}]
  &= - && e^{- i \theta} \tanh\chi \sec\zeta \, \tensor{v}_{*}
  \, ,
\end{alignat}
}

\section{Verma Modules of Smooth Vector Fields, 1-Forms and/or 3-Forms}\label{sect:Verma-modules}

As we have seen in the above, for an irreducible $\mathfrak{so}(1, 4)$ module
of smooth vector fields on $dS^{4}$, its highest weight $\lambda$ is either
$N \lambda_{1}$ or $N \lambda_{1} + 2 \lambda_{2}$, with $N$ a non-negative
integer.  Given such a highest weight $\lambda$, the corresponding Verma module
of smooth vector fields on $dS^{4}$ will be denoted by
$\mathfrak{X}(dS^{4})_{\lambda}$, which is spanned by the following
vector fields:
\begin{equation}
  \tensor{v}_{\lambda}^{(jklm)}
  := \Lied{\tensor{f}_{\alpha_{1} + \alpha_{2}}}^{j}
    \Lied{\tensor{f}_{\alpha_{1} + 2 \alpha_{2}}}^{k}
    \Lied{\tensor{f}_{\alpha_{1}}}^{l}
    \Lied{\tensor{f}_{\alpha_{2}}}^{m}
      \tensor{v}_{\lambda}
  \label{eq:v-lambda:jklm}
\end{equation}
for some non-negative integer $j$, $k$, $l$ and $m$.  Here, the notation
$\Lied{\tensor{f}_{\alpha_{2}}}^{j}$ stands for the action of the Lie derivative
$\Lied{\tensor{f}_{\alpha_{2}}}$ for $j$ times (provided that $j > 0$), while
$\Lied{\tensor{f}_{\alpha_{2}}}^{0}$ stands for the identity map, and so on.
In this manner each $\tensor{v} \in \mathfrak{X}(dS^{4})_{\lambda}$
can be written as a linear combination
\begin{equation}
  \tensor{v}
  = \sum_{j = 0}^{j_{\mathrm{max}}}
    \sum_{k = 0}^{k_{\mathrm{max}}}
    \sum_{l = 0}^{l_{\mathrm{max}}}
    \sum_{m = 0}^{m_{\mathrm{max}}}
    C_{jklm} \tensor{v}_{\lambda}^{(jklm)}
\end{equation}
for some constants $C_{jklm}$, where the integers $j_{\mathrm{max}}$,
$k_{\mathrm{max}}$, $l_{\mathrm{max}}$ and $m_{\mathrm{max}}$ can be determined
by the knowledge of the weight diagram. This will be described later.

Via the Lie derivatives, the Lie algebra $\mathfrak{so}(1, 4)$ also acts on
$\Omega^{p}(dS^{4})$, the infinite dimensional vector space of $p$-forms
on $dS^{4}$.
Given a dominant weight $\lambda = N_{1} \lambda_{1} + N_{2} \lambda_{2}$,
there might be a $p$-form $\vect{\alpha}_{\lambda}$ satisfying
\begin{alignat}{2}
  \Lied{\tensor{h}_{\alpha_{1}}} \vect{\alpha}_{\lambda}
  &= N_{1} \vect{\alpha}_{\lambda}
  \, , \qquad &
  \Lied{\tensor{h}_{\alpha_{2}}} \vect{\alpha}_{\lambda}
  &= N_{2} \vect{\alpha}_{\lambda}
  \, ,
  \label{eq:Lie-der:alpha-lambda:h}
  \\
  \Lied{\tensor{e}_{\alpha_{1}}} \vect{\alpha}_{\lambda} &= 0
  \, , &
  \Lied{\tensor{e}_{\alpha_{1} + 2 \alpha_{2}}} \vect{\alpha}_{\lambda} &= 0
  \, , \\
  \Lied{\tensor{e}_{\alpha_{2}}} \vect{\alpha}_{\lambda} &= 0
  \, , &
  \Lied{\tensor{e}_{\alpha_{1} + \alpha_{2}}} \vect{\alpha}_{\lambda} &= 0
  \, .
  \label{eq:Lie-der:alpha-lambda:e2}
\end{alignat}
Were there such a nonzero $p$-form $\vect{\alpha}_{\lambda}$, there will be
a corresponding Verma module $\Omega^{p}(dS^{4})_{\lambda}$ with the highest
weight $\lambda$, spanned by the $p$-forms such like
\begin{equation}
  \vect{\alpha}_{\lambda}^{(jklm)}
  := \Lied{\tensor{f}_{\alpha_{1} + \alpha_{2}}}^{j}
    \Lied{\tensor{f}_{\alpha_{1} + 2 \alpha_{2}}}^{k}
    \Lied{\tensor{f}_{\alpha_{1}}}^{l}
    \Lied{\tensor{f}_{\alpha_{2}}}^{m}
      \vect{\alpha}_{\lambda}
  \, ,
\end{equation}
hence a $p$-form $\vect{\alpha} \in \Omega^{p}(dS^{4})_{\lambda}$ can be
expressed as
\begin{equation}
  \vect{\alpha}
  = \sum_{j = 0}^{j_{\mathrm{max}}}
    \sum_{k = 0}^{k_{\mathrm{max}}}
    \sum_{l = 0}^{l_{\mathrm{max}}}
    \sum_{m = 0}^{m_{\mathrm{max}}}
    C_{jklm} \vect{\alpha}_{\lambda}^{(jklm)}
\end{equation}
for some constants $C_{jklm}$.  Similarly, the integers $j_{\mathrm{max}}$,
$k_{\mathrm{max}}$, $l_{\mathrm{max}}$ and $m_{\mathrm{max}}$ can be determined
by the knowledge of the weight diagram.

For $p = 0$, $\Omega^{0}(dS^{4})$ and $\Omega^{0}(dS^{4})_{\lambda}$
are identified with $C^{\infty}(dS^{4})$ and $C^{\infty}(dS^{4})_{\lambda}$,
respectively. According to the study in \cite{Zhou:2011dr}, possible $\lambda$ for the existence of $C^{\infty}(dS^{4})_{\lambda}$ could be merely
$\lambda = N \lambda_{1}$.

Since the Hodge $*$-operator is determined by the orientation and the metric
on $dS^{4}$,  it is straightforward that the Lie derivative with respect to
a Killing vector field $\tensor{X}$ is commutative with $*$:
\begin{equation}
  * \Lied{\tensor{X}} = \Lied{\tensor{X}} *
\end{equation}
As a consequence, a $p$-form $\vect{\alpha}_{\lambda}$
satisfying eqs.~(\ref{eq:Lie-der:alpha-lambda:h}) to
(\ref{eq:Lie-der:alpha-lambda:e2}) generates a $(4 - p)$-form
\begin{equation}
  \vect{\alpha}'_{\lambda} := *\vect{\alpha}_{\lambda}
  \, ,
\end{equation}
which also satisfies a set of equations similar to
eqs.~(\ref{eq:Lie-der:alpha-lambda:h}) to (\ref{eq:Lie-der:alpha-lambda:e2}).
Therefore, when the Verma module $\Omega^{p}(dS^{4})_{\lambda}$ exists,
there is also a Verma module
$\Omega^{4 - p}(dS^{4})_{\lambda} = *\Omega^{p}(dS^{4})_{\lambda}$
with the same highest weight $\lambda$.  What's more, these two Verma modules
are isomorphic to each other, which can be seen from
\begin{equation}
  {\vect{\alpha}'}^{(jklm)}_{\lambda} = * \vect{\alpha}^{(jklm)}_{\lambda}
  \, .
\end{equation}
Here $\vect{\alpha}'^{(jklm)}_{\lambda}$ is defined by
\begin{equation}
  {\vect{\alpha}'}^{(jklm)}_{\lambda}
  := \Lied{\tensor{f}_{\alpha_{1} + \alpha_{2}}}^{j}
    \Lied{\tensor{f}_{\alpha_{1} + 2 \alpha_{2}}}^{k}
    \Lied{\tensor{f}_{\alpha_{1}}}^{l}
    \Lied{\tensor{f}_{\alpha_{2}}}^{m}
      \vect{\alpha}_{\lambda}'
  \, .
\end{equation}

For convenience, the map sending a vector field $\tensor{v}$ to a 1-form
$\tilde{\tensor{v}}$, or $v^{a}$ to $v_{a} = g_{ab} v^{b}$ in the convention of
abstract indices, is denoted by
$\tensor{g}_{\flat} \colon \mathfrak{X}(dS^{4}) \rightarrow \Omega^{1}(dS^{4})$
in this paper.  Then, for each Killing vector field $\tensor{X}$,
there is $\Lied{\tensor{X}} v_{a} = g_{ab} \, \Lied{\tensor{X}} v^{b}$
for arbitrary vector field $v^{a}$, namely,
\begin{equation}
  \Lied{\tensor{X}} \tensor{g}_{\flat} = \tensor{g}_{\flat} \Lied{\tensor{X}}
  \, .
\end{equation}
This indicates that $\tensor{g}_{\flat}$ is an isomorphism of
$\mathfrak{so}(1, 4)$-modules:  with respect to the actions of
$\mathfrak{so}(1, 4)$, each property of $\mathfrak{X}(dS^{4})$ is precisely
mapped to $\Omega^{1}(dS^{4})$.  Especially, the Verma module
$\mathfrak{X}(dS^{4})_{\lambda}$ is mapped by $\tensor{g}_{\flat}$ to
the Verma module $\Omega^{1}(dS^{4})_{\lambda}$.  As a consequence, possible
highest weights for $\Omega^{1}(dS^{4})_{\lambda}$ are either $N \lambda_{1}$
or $N \lambda_{1} + 2 \lambda_{2}$, with $N$ a non-negative integer.
Then, by the Hodge $*$-operator, these are also the highest weights of
$\Omega^{3}(dS^{4})_{\lambda}$.

For the highest weight $\lambda = N \lambda_{1}$, it is the most convenient
to investigate $\Omega^{1}(dS^{4})_{N \lambda_{1}}$, instead of
$\mathfrak{X}(dS^{4})_{N \lambda_{1}}$ and $\Omega^{3}(dS^{4})_{N \lambda_{1}}$,
because the highest weight 1-form
\begin{equation}
  \tilde{\tensor{v}}_{N \lambda_{1}}
  = \tensor{g}_{\flat} \tensor{v}_{N \lambda_{1}}
  = l \, \od \phi_{N \lambda_{1}}
  \, ,
\end{equation}
as shown in eq.~(\ref{gab-vb:N-lambda1}).  Then, due to
$\Lied{\tensor{X}} \od = \od \Lied{\tensor{X}}$ for an arbitrary vector field
$\tensor{X}$,
\begin{alignat*}{1}
  \tilde{\tensor{v}}^{(jklm)}_{N \lambda_{1}}
  &= \Lied{\tensor{f}_{\alpha_{1} + \alpha_{2}}}^{j}
    \Lied{\tensor{f}_{\alpha_{1} + 2 \alpha_{2}}}^{k}
    \Lied{\tensor{f}_{\alpha_{1}}}^{l}
    \Lied{\tensor{f}_{\alpha_{2}}}^{m}
    \tilde{\tensor{v}}_{N \lambda_{1}}
  = l \, \od (
      \Lied{\tensor{f}_{\alpha_{1} + \alpha_{2}}}^{j}
      \Lied{\tensor{f}_{\alpha_{1} + 2 \alpha_{2}}}^{k}
      \Lied{\tensor{f}_{\alpha_{1}}}^{l}
      \Lied{\tensor{f}_{\alpha_{2}}}^{m}
      \phi_{N \lambda_{1}}
    )
  \, .
\end{alignat*}
Because of $\Lied{\tensor{f}_{\alpha_{2}}} \phi_{N \lambda_{1}} = 0$ (see, \cite{Zhou:2011dr}), there is
$\tilde{\tensor{v}}^{(jklm)}_{N \lambda_{1}} = 0$ whenever $m > 0$.
So, nonzero $\tilde{\tensor{v}}^{(jklm)}_{N \lambda_{1}}$ is among those
with $m = 0$:
\begin{equation}
  \tilde{\tensor{v}}^{(jkl0)}_{N \lambda_{1}}
  = l \, \od \phi^{(jkl)}_{N \lambda_{1}}
  \, .
\end{equation}
For the expression of $\phi^{(jkl)}_{N \lambda_{1}}$, we refer to \cite{Zhou:2011dr}. Thus the Verma module
\begin{equation}
  \Omega^{1}(dS^{4})_{N \lambda_{1}}
  = \od [\Omega^{0}(dS^{4})_{N \lambda_{1}}]
  = \od [ C^{\infty}(dS^{4})_{N \lambda_{1}} ]
  \, ,
\end{equation}
consisting of exact 1-forms only.

The Verma modules $\mathfrak{X}(dS^{4})_{N \lambda_{1}}$,
$\Omega^{1}(dS^{4})_{N \lambda_{1}}$ and $\Omega^{3}(dS^{4})_{N \lambda_{1}}$
are all irreducible $\mathfrak{so}(1, 4) \otimes \mathbb{C}$-modules.

For the highest weight $\lambda = N \lambda_{1} + 2 \lambda_{2}$, it is
the most convenient to investigate $\mathfrak{X}(dS^{4})_{\lambda}$, instead of
$\Omega^{1}(dS^{4})_{\lambda}$ and $\Omega^{3}(dS^{4})_{\lambda}$.
This module is spanned by vector fields like those
in eq.~(\ref{eq:v-lambda:jklm}), for which $\tensor{v}_{\lambda}$ is as shown
in eq.~(\ref{eq:v-lambda}).
Their expressions are listed below:
\begin{alignat}{1}
  \tensor{v}_{\lambda}^{(jk \ell 0)}
  &= \frac{2}{l} \phi_{N \lambda_{1}}^{(jk \ell)} \,
    \tensor{e}_{\alpha_{1} + 2 \alpha_{2}}
  - \frac{2 j}{l} \phi_{N \lambda_{1}}^{(j - 1, k \ell)} \,
    \tensor{e}_{\alpha_{2}}
  - \frac{2 k}{l} \phi_{N \lambda_{1}}^{(j, k - 1, \ell)} \,
    (\tensor{h}_{\alpha_{1}} + \tensor{h}_{\alpha_{2}})
  + \frac{2 j (j - 1)}{l} \phi_{N \lambda_{1}}^{(j - 2, k \ell)} \,
    \tensor{f}_{\alpha_{1}}
  \nonumber \\ & \phantom{={}}
  - \frac{2 jk}{l} \phi_{N \lambda_{1}}^{(j - 1, k - 1, \ell)} \,
    \tensor{f}_{\alpha_{1} + \alpha_{2}}
  - \frac{2 k \, (k - 1)}{l} \phi_{N \lambda_{1}}^{(j, k - 2, \ell)} \,
    \tensor{f}_{\alpha_{1} + 2 \alpha_{2}}
  \, ,
  \\
  \tensor{v}_{\lambda}^{(jkl1)}
  &= \frac{2}{l} \, \phi_{N \lambda_{1}}^{(jk \ell)} \,
    \tensor{e}_{\alpha_{1} + \alpha_{2}}
  + \frac{2 \ell}{l} \, \phi_{N \lambda_{1}}^{(jk, \ell - 1)} \,
    \tensor{e}_{\alpha_{2}}
  - \frac{2j}{l} \, \phi_{N \lambda_{1}}^{(j - 1, k \ell)} \,
    (2 \, \tensor{h}_{\alpha_{1}} + \tensor{h}_{\alpha_{2}})
  - \frac{4j \ell}{l} \, \phi_{N \lambda_{1}}^{(j - 1, k, \ell - 1)} \,
    \tensor{f}_{\alpha_{1}}
  - \frac{2 k}{l} \, \phi_{N \lambda_{1}}^{(j, k - 1, \ell)} \,
    \tensor{f}_{\alpha_{2}}
  \nonumber \\ & \phantom{={}}
  + \frac{2}{l} \, (
      k \ell \, \phi_{N \lambda_{1}}^{(j, k - 1, \ell - 1)}
      - j(j - 1) \, \phi_{N \lambda_{1}}^{(j - 2, k \ell)}
    ) \,
    \tensor{f}_{\alpha_{1} + \alpha_{2}}
  - \frac{4 jk}{l} \, \phi_{N \lambda_{1}}^{(j - 1, k - 1, \ell)} \,
    \tensor{f}_{\alpha_{1} + 2 \alpha_{2}}
  \, .
  \\
  \tensor{v}_{\lambda}^{(jk \ell 2)}
  &=  - \frac{4}{l} \, \phi_{N \lambda_{1}}^{(jk \ell)} \,
    \tensor{e}_{\alpha_{1}}
  + \frac{4 \ell}{l} \, \phi_{N \lambda_{1}}^{(jk, \ell - 1)} \,
    \tensor{h}_{\alpha_{1}}
  + \frac{4 \ell (\ell - 1)}{l} \, \phi_{N \lambda_{1}}^{(jk, \ell - 2)} \,
    \tensor{f}_{\alpha_{1}}
  - \frac{4j}{l} \, \phi_{N \lambda_{1}}^{(j - 1, k \ell)} \,
    \tensor{f}_{\alpha_{2}}
  \nonumber \\ & \phantom{={}}
  + \frac{4j \ell}{l} \, \phi_{N \lambda_{1}}^{(j - 1, k, \ell - 1)} \,
    \tensor{f}_{\alpha_{1} + \alpha_{2}}
  - \frac{4j (j - 1)}{l} \, \phi_{N \lambda_{1}}^{(j - 2, k \ell)} \,
    \tensor{f}_{\alpha_{1} + 2 \alpha_{2}}
  \, ,
  \\
  \tensor{v}_{\lambda}^{(jk \ell m)} &= 0
  \, , \qquad (m > 2)
  \, .
\end{alignat}

\section{Properties of 1-Forms in Each Irreducible $\mathfrak{so}(1, 4)$-Module}\label{sect:properties}

For a vector field $\tensor{v}$, the 1-form $v_{a} = g_{ab} v^{b}$ is also
denoted by $\tilde{\tensor{v}}$ without abstract indices, in this paper.
For a dominant weight $\lambda = N \lambda_{1} + 2 \lambda_{2}$,
the corresponding highest weight vector field $\tensor{v}_{\lambda}$,
as shown in eq.(\ref{eq:v-lambda}), has
\begin{equation}
  \tilde{\tensor{v}}_{\lambda}
  = \frac{2}{l} \phi_{N \lambda_{1}} \,
    \tilde{\tensor{e}}_{\alpha_{1} + 2 \alpha_{2}}
  = -l \, (\cosh\chi)^{N + 2} (\cos\zeta)^{N} e^{-i (N + 1) \theta - i \varphi}
    [\od\zeta + i \sin\zeta \cos\zeta \, (\od\theta - \od\varphi)]
  \, .
\end{equation}
It is straightforward to obtain that
\begin{alignat}{1}
  \od\tilde{\tensor{v}}_{\lambda}
  &= - (N + 2) l \phi_{\lambda_{1}}^{N} e^{- i \theta - i \varphi}
    \sinh\chi \cosh\chi \; \od\chi \wedge
    [ \od\zeta + i \sin\zeta \cos\zeta \, (\od\theta - \od\varphi) ]
  \nonumber \\ & \phantom{={}}
  - (N + 2) l \phi_{\lambda_{1}}^{N} e^{- i \theta - i \varphi}
    \cosh^{2} \chi \, (
      i \cos^{2} \zeta \, \od\zeta \wedge \od\theta
      + i \sin^{2} \zeta \, \od\zeta \wedge \od\varphi
      - \sin\zeta \cos\zeta \, \od\theta \wedge \od\varphi
    )
  \, .
\end{alignat}

The coordinate system $(\chi, \zeta, \theta, \varphi)$ is not always compatible
with the orientation of $dS^{4}$.  That means, the volume 4-form is not
necessarily
$\sqrt{|g|} \, \od\chi \wedge \od\zeta \wedge \od\theta \wedge \od\varphi$:
sometimes it could differ by a negative sign.  To obtain the correct expression
of the volume 4-form $\vect{\varepsilon}$ on $dS^{4}$, we notice that
on $\mathbb{R}^{1, 4}$ there is the standard volume 5-form
$\overline{\vect{\varepsilon}} = \od \xi^{0} \wedge \od \xi^{1} \wedge \cdots
  \wedge \od \xi^{4}$.
Then, setting
\begin{equation}
  \tensor{D} = \xi^{A} \frac{\partial}{\partial \xi^{A}}
  \, ,
\end{equation}
the volume 4-form $\vect{\varepsilon}$ on $dS^{4}$ is just the pull-back of
$\frac{1}{l} \, i_{\tensor{D}} \overline{\vect{\varepsilon}}$
to the hypersurface $dS^{4}$ in $\mathbb{R}^{1, 4}$.
In terms of the coordinates $\chi$, $\zeta$, $\theta$ and $\varphi$,
\begin{equation}
  \vect{\varepsilon} = l^{4} \cosh^{3} \chi \sin\zeta \cos\zeta \;
    \od\chi \wedge \od\zeta \wedge \od\theta \wedge \od\varphi
  \, .
\end{equation}
For the Hodge dual,
\begin{alignat}{3}
  & *(\od\chi \wedge \od\zeta)
  &&= {} & - & \cosh\chi \sin\zeta \cos\zeta \, \od\theta \wedge \od\varphi
  \, , \\
  & *(\od\chi \wedge \od\theta)
  &&= & & \cosh\chi \tan\zeta \, \od\zeta \wedge \od\varphi
  \, , \\
  & *(\od\chi \wedge \od\varphi)
  &&= {} & - & \cosh\chi \cot\zeta \, \od\zeta \wedge \od\theta
  \, , \\
  & *(\od\zeta \wedge \od\theta)
  &&= & & \frac{\tan\zeta}{\cosh\chi} \, \od\chi \wedge \od\varphi
  \, , \\
  & *(\od\zeta \wedge \od\varphi)
  &&= {} & - & \frac{\cot\zeta}{\cosh\chi} \, \od\chi \wedge \od\theta
  \, , \\
  & *(\od\theta \wedge \od\varphi)
  &&= & & \frac{\sec\zeta \csc\zeta}{\cosh\chi} \, \od\chi \wedge \od\zeta
  \, .
\end{alignat}
By virtue of these relations, one has
\begin{alignat}{1}
  * \, \od * \od \tilde{\tensor{v}}_{\lambda}
  = - \frac{(N + 2)(N + 3)}{l^{2}} \, \tilde{\tensor{v}}_{\lambda}
  \, .
  \label{eq:*d*d:v-lambda}
\end{alignat}
Consequently,
\begin{equation}
  * \, \od * \tilde{\tensor{v}}_{\lambda} = 0
  \, .
  \label{eq:*d*:v-lambda}
\end{equation}

Since both $*$ and $\od$ commute with the Lie derivatives, one has
\begin{alignat}{1}
  * \, \od * \od \tilde{\tensor{v}}_{\lambda}^{(jk \ell m)}
  &= - \frac{(N + 2)(N + 3)}{l^{2}} \tilde{\tensor{v}}_{\lambda}^{(jk \ell m)}
  \, , \\
  * \, \od * \tilde{\tensor{v}}_{\lambda}^{(jk \ell m)} &= 0
  \, ,
\end{alignat}
for arbitrary nonnegative integers $j$, $k$, $\ell$ and $m$.
Hence every 1-form $\tilde{\tensor{v}} \in \Omega^{1}(dS^{4})_{\lambda}$,
with $\lambda = N \lambda_{1} + 2 \lambda_{2}$, satisfies
\begin{alignat}{1}
  * \, \od * \od \tilde{\tensor{v}}
  &= - \frac{(N + 2)(N + 3)}{l^{2}} \tilde{\tensor{v}}
  \, ,
  \label{eq:*d*d:v}
  \\
  * \, \od * \tilde{\tensor{v}} &= 0
  \, .
\end{alignat}

For a dominant weight $\lambda = N \lambda_{1}$, the corresponding
highest weight vector field $\tensor{v}_{N \lambda_{1}}$ has
\begin{equation}
  \tilde{\tensor{v}}_{N \lambda_{1}} = \od \, (l \phi_{N \lambda_{1}})
  = Nl \, \phi_{\lambda_{1}}^{N - 1} \, \od \phi_{\lambda_{1}}
  \, ,
\end{equation}
as indicated in eq.~(\ref{gab-vb:N-lambda1}).  Then, due to the identities
\begin{equation}
  - \nabla_{a} v_{N \lambda_{1}}^{a}
  = * \, \od * \tilde{\tensor{v}}_{N \lambda_{1}}
  = * \, \od * \od (l \phi_{N \lambda_{1}})
  = - g^{ab} \nabla_{a} \nabla_{b} (l \phi_{N \lambda_{1}})
\end{equation}
and the result in \cite{Zhou:2011dr}, there will be
\begin{equation}
  * \, \od * \tilde{\tensor{v}}_{N \lambda_{1}}
  = - \frac{N (N + 3)}{l} \phi_{N \lambda_{1}}
  = - \nabla_{a} v_{N \lambda_{1}}^{a}
  \, .
\end{equation}
Note that $\tilde{\tensor{v}}_{N \lambda_{1}}$ is zero when $N = 0$.
Hence the $\nabla_{a} v_{N \lambda_{1}}^{a}$ is nonzero whenever
$\tilde{\tensor{v}}_{N \lambda_{1}}$ is nonzero.
It follows that, for each
$\tilde{\tensor{v}} \in \Omega^{1}(dS^{4})_{N \lambda_{1}}$,
there is a smooth function $\phi$ on $dS^{4}$ satisfying
\begin{equation}
  \tilde{\tensor{v}} = \od \phi
  \, , \qquad
  * \, \od * \tilde{\tensor{v}}
  = - \nabla_{a} v^{a}
  = - \frac{N (N + 3)}{l} \phi
  \, .
\end{equation}
In fact, such a function $\phi$ could be chosen from the irreducible
$\mathfrak{so}(1, 4)$-module $C^{\infty}(dS^{4})_{N \lambda_{1}}$.

\begin{alignat}{1}
  g_{ab} v_{N \lambda_{1} + 2 \lambda_{2}}^{a}
    v_{N \lambda_{1} + 2 \lambda_{2}}^{b}
  &= \frac{4}{l^{2}} \, \phi_{2N \lambda_{1}}
    g_{ab} \, e_{\alpha_{1} + 2 \alpha_{2}}^{a}
    e_{\alpha_{1} + 2 \alpha_{2}}^{b}.
\end{alignat}

\section{Smooth Solutions of the Maxwell Equations on $dS^{4}$}\label{sect:smooth-solutions}

Now that we have obtained all finite dimensional irreducible
$\mathfrak{so}(1, 4)$-submodules of $\Omega^{1}(dS^{4})$,
the space of smooth 1-forms on $dS^{4}$, there is the direct sum decomposition
\begin{alignat}{1}
  \Omega^{1}(dS^{4})
  &= \Omega^{1}(dS^{4})_{\mathrm{fin}} \oplus \Omega^{1}(dS^{4})_{\infty}
  \, ,
  \\
  \Omega^{1}(dS^{4})_{\mathrm{fin}}
  & := \bigoplus_{N = 0}^{\infty} \Omega^{1}(dS^{4})_{N \lambda_{1}}
  \oplus \bigoplus_{N = 0}^{\infty}
    \Omega^{1}(dS^{4})_{N \lambda_{1} + 2 \lambda_{2}}
  \, .
\end{alignat}
Here $\Omega^{1}(dS^{4})_{\infty}$ is the direct of all infinite dimensional
irreducible $\mathfrak{so}(1, 4)$-submodules of $\Omega^{1}(dS^{4})$ or
just the zero subspace, depending on the existence of infinite dimensional
irreducible $\mathfrak{so}(1, 4)$-submodules in $\Omega^{1}(dS^{4})$.

It remains an open problem whether $\Omega^{1}(dS^{4})$ has
an infinite dimensional irreducible $\mathfrak{so}(1, 4)$-submodule.
Currently we discuss smooth solutions the Maxwell equations with the work
assumption that $\Omega^{1}(dS^{4})_{\infty} = 0$.  In other words,
when we talk about a smooth 1-form (or, equivalently, a smooth vector field),
we are referring to one contained in the direct sum
$\Omega^{1}(dS^{4})_{\mathrm{fin}}$.

Thus a 4-potential $\tensor{A} = A_{\mu} \, \od x^{\mu}$ can be uniquely
decomposed into
\begin{equation}
  \tensor{A}
  = \sum_{N}
    ( \tensor{A}_{N \lambda_{1}} + \tensor{A}_{N \lambda_{1} + 2 \lambda_{2}})
  \label{eq:A:decomposition:1}
\end{equation}
with $\tensor{A}_{N \lambda_{1}} \in \Omega^{1}(dS^{4})_{N \lambda_{1}}$
and $\tensor{A}_{N \lambda_{1} + 2 \lambda_{2}} \in
  \Omega^{1}(dS^{4})_{N \lambda_{1} + 2 \lambda_{2}}$.
If one requires the 4-potential $\tensor{A}$ to satisfy
the Lorentz gauge condition
\begin{equation}
  - \nabla_{a} A^{a} = * \, \od * \tensor{A} = 0
  \, ,
\end{equation}
all $\tensor{A}_{N \lambda_{1}}$ must be zero.  Hence one has a unique
decomposition
\begin{equation}
  \tensor{A} = \sum_{N} \tensor{A}_{N \lambda_{1} + 2 \lambda_{2}}
  \label{eq:A:decomposition}
\end{equation}
subject to the Lorentz gauge condition.

Given an electric/magnetic 4-current $\tensor{J}$,  the continuity equation
\begin{equation}
  - \nabla_{a} J^{a} = * \, \od * \tilde{\tensor{J}} = 0
\end{equation}
implies that there is the unique decomposition
\begin{equation}
  \tilde{\tensor{J}}
  = \sum_{N} \tilde{\tensor{J}}_{N \lambda_{1} + 2 \lambda_{2}}
  \, ,
\end{equation}
where $\tilde{\tensor{J}}_{N \lambda_{1} + 2 \lambda_{2}}
  \in \Omega^{1}(dS^{4})_{N \lambda_{1} + 2 \lambda_{2}}$.

Now consider the solution
\begin{equation}
  \tensor{F}^{\textrm{(e)}} = \od \tensor{A}^{\mathrm{(e)}}
\end{equation}
of the Maxwell equations
\begin{equation}
  \od \tensor{F}^{\textrm{(e)}} = 0
  \, , \quad
  * \, \od * \tensor{F}^{\textrm{(e)}}
  = - \frac{4 \pi}{c} \tilde{\tensor{J}}^{\textrm{(e)}}
  \label{eq:Maxwell:e}
\end{equation}
in Gaussian units.  Having got the unique decompositions
\begin{equation}
  \tensor{J}^{\textrm{(e)}}
  = \sum_{N} \tensor{J}^{\mathrm{(e)}}_{N \lambda_{1} + 2 \lambda_{2}}
  \, , \quad
  \tensor{A}^{\textrm{(e)}}
  = \sum_{N} \tensor{A}^{\mathrm{(e)}}_{N \lambda_{1} + 2 \lambda_{2}}
  \, ,
\end{equation}
the Maxwell equations and eq.~(\ref{eq:*d*d:v}) result in
\begin{alignat}{1}
  \tilde{\tensor{J}}^{\mathrm{(e)}}_{N \lambda_{1} + 2 \lambda_{2}}
  &= \frac{c}{4 \pi} \frac{(N + 2)(N + 3)}{l^{2}}
    \tensor{A}^{\mathrm{(e)}}_{N \lambda_{1} + 2 \lambda_{2}}
  \, .
  \label{eq:J-A:e}
\end{alignat}

As for the Maxwell equations with magnetic 4-currents
$\tensor{J}^{\mathrm{(m)}}$ only, one has
\begin{equation}
  \od * \tensor{F}^{\textrm{(m)}} = 0
  \, , \qquad
  * \, \od \tensor{F}^{\textrm{(m)}}
  = - \frac{4 \pi}{c} \tilde{\tensor{J}}^{\textrm{(m)}}
  \, .
  \label{eq:Maxwell:m}
\end{equation}
Comparing them with eqs.~(\ref{eq:Maxwell:e}), the solution is rather simple:
\begin{equation}
  \tensor{F}^{\textrm{(m)}} = * \, \od \tensor{A}^{\mathrm{(m)}}
\end{equation}
with $\tensor{A}^{\mathrm{(m)}}$ the similar decomposition as $\tensor{A}$
in eq.~(\ref{eq:A:decomposition}), provided that the Lorentz gauge condition
is also required.  Then the Maxwell equations require that
\begin{equation}
  \tensor{J}^{\textrm{(m)}}
  = \sum_{N} \tensor{J}^{\mathrm{(m)}}_{N \lambda_{1} + 2 \lambda_{2}}
  \, , \quad
  \tensor{A}^{\textrm{(m)}}
  = \sum_{N} \tensor{A}^{\mathrm{(m)}}_{N \lambda_{1} + 2 \lambda_{2}}
\end{equation}
satisfy
\begin{equation}
  \tilde{\tensor{J}}^{\textrm{(m)}}_{N \lambda_{1} + 2 \lambda_{2}}
  = \frac{c}{4 \pi} \frac{(N + 2)(N + 3)}{l^{2}}
    \tensor{A}^{\mathrm{(m)}}_{N \lambda_{1} + 2 \lambda_{2}}
  \, .
  \label{eq:J-A:m}
\end{equation}

For the Maxwell equations
\begin{equation}
  * \, \od \tensor{F} = - \frac{4 \pi}{c} \tilde{\tensor{J}}_{\textrm{m}}
  \, , \qquad
  * \, \od * \tensor{F} = - \frac{4 \pi}{c} \tilde{\tensor{J}}_{\textrm{e}}
  \label{eq:Maxwell:em}
\end{equation}
with both electric 4-current $\tensor{J}^{\mathrm{(e)}}$ and magnetic 4-current
$\tensor{J}^{\mathrm{(m)}}$, we can introduce two potentials
$\tensor{A}^{\mathrm{(e)}}$ and $\tensor{A}^{\mathrm{(m)}}$ so that
\begin{equation}
  \tensor{F} = * \, \od \tensor{A}^{\mathrm{(m)}}
  + \od \tensor{A}^{\mathrm{(e)}}
  \, .
  \label{eq:F:em}
\end{equation}
Then the Maxwell equations force the relations (\ref{eq:J-A:e}) and
(\ref{eq:J-A:m}) to be satisfied, provided that the Lorentz gauge condition
is satisfied by these potentials, respectively.

So far we have discussed the Maxwell equations by virtue of guage potentials,
under the Lorentz gauge condition.
By virtue of the direct sum decomposition for both 4-potentials and 4-currents,
$\tensor{A}^{\mathrm{(e)}}$  (or $\tensor{A}^{\mathrm{(m)}}$) and
$\tensor{J}^{\mathrm{(e)}}$  (or $\tensor{J}^{\mathrm{(m)}}$) can be mutually
determined, as shown in eq.(\ref{eq:J-A:e}) (or eq.~(\ref{eq:J-A:m})).
It is concluded that there will be no source-free smooth electromagnetic fields
in $dS^{4}$:  if there is neither charge nor current, no matter electric or
magnetic, eqs.(\ref{eq:J-A:e}) and (\ref{eq:J-A:m})
will force the corresponding electric/magnetic 4-potential to be zero,
which results in a zero electromagnetic field, according to eq.(\ref{eq:F:em}).

\section{The Mass of Proca Equation}\label{sect:mass-Proc}

Subject to the Lorentz gauge condition
\begin{equation}
  \nabla^{a} A^{\mathrm{(e)}}_{a} = 0
  \, , \qquad
  \nabla^{a} A^{\mathrm{(m)}}_{a} = 0
\end{equation}
as well as eq.(\ref{eq:F:em}), the Maxwell equations (\ref{eq:Maxwell:em})
have the following  equivalent form:
\begin{equation}
  \nabla_{b} \nabla^{b} A^{\mathrm{(e)}}_{a}
  - R_{a}^{\phantom{a} b} A^{\mathrm{(e)}}_{b}
  = \frac{4 \pi}{c} J^{\textrm{(e)}}_{a}
  \, , \qquad
  \nabla_{b} \nabla^{b} A^{\mathrm{(m)}}_{a}
  - R_{a}^{\phantom{a} b} A^{\mathrm{(m)}}_{b}
  = \frac{4 \pi}{c} J^{\textrm{(m)}}_{a}
  \, .
\end{equation}
The Ricci tensor of $dS^{4}$ is $R_{ab} = - \frac{3}{l^{2}} g_{ab}$.
(The negative sign is due to the convention for the metric signature.)
Substituting this and eqs.(\ref{eq:J-A:e}), (\ref{eq:J-A:m}) into the above
equation, one has
\begin{equation}
  \nabla_{b} \nabla^{b} A_{N \lambda_{1} + 2 \lambda_{2}}^{a}
  - \frac{N^{2} + 5 N + 3}{l^{2}} A_{N \lambda_{1} + 2 \lambda_{2}}^{a}
  = 0
\end{equation}
for both $\tensor{A}^{\mathrm{(e)}}_{N \lambda_{1} + 2 \lambda_{2}}$
and $\tensor{A}^{\mathrm{(m)}}_{N \lambda_{1} + 2 \lambda_{2}}$,
for each $N \geqslant 0$.  This is a Proca equation with an imaginary mass:
\begin{equation}
  m^{2} = - \frac{N (N + 5) + 3}{l^{2}}
  \, .
\end{equation}


\section{Conclusions and Discussion}\label{sect:CD}

Applying the theory of Lie groups and Lie algebras, we have obtained all
finite dimensional irreducible $\mathfrak{so}(1, 4)$-submodules of
$\Omega^{1}(dS^{4})$, the space of differential 1-forms on $dS^{4}$.
All such $\mathfrak{so}(1,4)$-submodules are Verma modules
$\Omega^{1}(dS^{4})_{\lambda}$,
with the highest weight $\lambda$ being either $N \lambda_{1}$ or
$N \lambda_{1} + 2 \lambda_{2}$, where $N$ is a nonnegative integer.

It remains an open problem whether there is an infinite dimensional
irreducible $\mathfrak{so}(1, 4)$-submodule of $\Omega^{1}(dS^{4})$.
If there is no such an $\mathfrak{so}(1, 4)$-submodule in $\Omega^{1}(dS^{4})$,
every smooth 1-form on $dS^{4}$, such as $\tensor{A}$ say, can be decomposed uniquely
into a sum as shown in eq.~(\ref{eq:A:decomposition:1}).
Otherwise an additional term
$\tensor{A}_{\infty} \in \Omega^{1}(dS^{4})_{\infty}$ is needed on the right
hand side of eq.(\ref{eq:A:decomposition:1}).

As $\mathfrak{so}(1, 4)$-modules, $\mathfrak{X}(dS^{4})$ (the space of smooth
vector fields on $dS^{4}$) and $\Omega^{3}(dS^{4})$ are isomorphic to
$\Omega^{1}(dS^{4})$.  Hence the structures of these three spaces are totally
similar.

If $\Omega^{1}(dS^{4})_{\infty} = 0$, the smooth solution of
the Maxwell equations can be directly determined by the smooth electric/magnetic
4-current(s), without neither initial nor boundary conditions.
This is as shown in eqs.(\ref{eq:J-A:e}) and (\ref{eq:J-A:m}).
An important consequence is that, on $dS^{4}$, there is no smooth source-free
electromagnetic fields.

Perhaps one would argue that there might be pointed (electric or magnetic)
charges so that its electromagnetic fields are still source-free
off the position of the charges. However, such electromagnetic fields, if exist,
is no longer smooth everywhere in the spacetime; and, more importantly,
such electromagnetic fields are not source-free, strictly speaking.
This problem will be left to discuss in other papers.

Electrodynamics in the Minkowski background tells us that an electromagnetic
wave propagates in vacuum at the speed $c$, carrying together stress tensor.  
This is why we accept that electromagnetic field is also
a physical existence, which does not depend on the existence of other matter.
Now we have shown that in the de~Sitter background it is not the case, and
the nonexistence of source-free electromagnetic fields means that
electromagnetic fields cannot exist independently.

Always companioned by source currents, electromagnetic fields cannot propagate
in vacuum in the de~Sitter background.  Our knowledge of electromagnetism
in the Minkowski background raises a question:  does the propagation of
electromagnetic wave (light) in $dS^{4}$ still satisfies the principle of
the invariance of light speed? This is an important question, concerning with
the foundation of measurement of the spacetime metric.

Before answering this question, involved concepts must be interpreted.
All these and the question itself will be left as a subject in other papers.
Here we only present some negative evidence.

The first evidence is that electromagnetic waves often propagate at a lower
speed than $c$ in a medium, according to electrodynamics in the Minkowski
background.  Now that electromagnetic fields cannot exist in vacuum (in the
classical sense), it could be strongly suspected that the propagation speed
of electromagnetic fields in $dS^{4}$ cannot be the ``light speed in vacuum''.

The second evidence is the potential-current relationship, namely,
eqs.(\ref{eq:J-A:e}) and (\ref{eq:J-A:m}):  if the electromagnetic wave
described by $\tensor{A}^{\mathrm{(e)}}$ or $\tensor{A}^{\mathrm{(m)}}$
``propagates at the light speed in vacuum'', then so does the source
described by $\tensor{J}^{\mathrm{(e)}}$ or $\tensor{J}^{\mathrm{(m)}}$.
However, our knowledge of the special relativity and quantum field theory tells
us that charged particles should not propagate like photons.  Although this is
based on the Minkowski background, cannot we raise a question if it is violated
in the de~Sitter background?

The third evidence is that the 4-potential
$\tensor{A}_{N \lambda_{1} + 2 \lambda_{2}}
  \in \Omega^{1}(dS^{4})_{N \lambda_{1} + 2 \lambda_{2}}$,
no matter electric or magnetic, satisfies the Proca equation with a mass
such that $m^{2} < 0$.  Our knowledge based on the Minkowski background reminds
us that $\tensor{A}_{N \lambda_{1} + 2 \lambda_{2}}$ cannot propagate
``at the light speed in vacuum''.

So far we just present a rough description of our questions, raising by
the electromagnetism in the de~Sitter background.  As have stated, to answer
these questions, we should first interpret the precise meaning of these
questions.  Both the meaning and the answer of these questions concerns
seriously with the foundation of physics, especially the foundation of
the measurement of spacetime distance in general relativity.  Foundation of
quantum field theory in curved spacetimes is also concerned with.  All these
questions will be investigated in further paper.

The conclusions for electromagnetic fields are heavily based on the 4-potentials
$\tensor{A}^{\mathrm{(e)}}$ and $\tensor{A}^{\mathrm{(m)}}$.
In fact, we can discuss the Maxwell equations without referring to 4-potentials,
while the conclusion remains correct.  This will be demonstrated in
forthcoming papers.

\section*{\centering Acknowledgements}

We thank Zhen-Hua Zhou, Wei Zhang, and Chun-Liang Liu for their helpful advice.

\appendix


\section{Connection 1-Forms and Curvature 2-Forms}

With respect to the coframe
\begin{alignat}{2}
\vect{\theta}^{0}= l\chi, \quad \vect{\theta}^{1}= l\cosh\chi\od\zeta,\quad \vect{\theta}^{2} = l \cosh\chi \cos\zeta \od\theta,
\quad \vect{\theta}^{3}= l \cosh\chi \sin\zeta\od\varphi,
\end{alignat}
the connection 1-forms are
\begin{alignat}{4}
  \vect{\omega}^{0}_{0} &= 0
  \, , &
  \vect{\omega}^{0}_{1} &= \frac{1}{l} \tanh\chi \, \vect{\theta}^{1}
  \, , &
  \vect{\omega}^{0}_{2} &= \frac{1}{l} \tanh\chi \, \vect{\theta}^{2}
  \, , \qquad &
  \vect{\omega}^{0}_{3} &= \frac{1}{l} \tanh\chi \, \vect{\theta}^{3}
  \, , \\
  \vect{\omega}^{1}_{0} &= \frac{1}{l} \tanh\chi \, \vect{\theta}^{1}
  \, , \qquad &
  \vect{\omega}^{1}_{1} &= 0
  \, , &
  \vect{\omega}^{1}_{2} &= \frac{\tan\zeta}{l \cosh\chi} \vect{\theta}^{2}
  \, , &
  \vect{\omega}^{1}_{3} &= - \frac{\cot\zeta}{l \cosh\chi} \vect{\theta}^{3}
  \, , \\
  \vect{\omega}^{2}_{0} &= \frac{1}{l} \tanh\chi \, \vect{\theta}^{2}
  \, , &
  \vect{\omega}^{2}_{1} &= - \frac{\tan\zeta}{l \cosh\chi} \vect{\theta}^{2}
  \, , \qquad &
  \vect{\omega}^{2}_{2} &= 0
  \, , &
  \vect{\omega}^{2}_{3} &= 0
  \, , \\
  \vect{\omega}^{3}_{0} &= \frac{1}{l} \tanh\chi \, \vect{\theta}^{3}
  \, , &
  \vect{\omega}^{3}_{1} &= \frac{\cot\zeta}{l \cosh\chi} \vect{\theta}^{3}
  \, , &
  \vect{\omega}^{3}_{2} &= 0
  \, , &
  \vect{\omega}^{3}_{3} &= 0
  \, .
\end{alignat}
As for the curvature 2-forms $\vect{\Omega}^{\mu}_{\nu}$,
\begin{alignat}{2}
  & \vect{\Omega}^{\mu}_{\nu}
  = \frac{1}{l^{2}} \vect{\theta}^{\mu} \wedge \vect{\theta}^{\nu}
  \, , & \mbox{if}\quad  \mu \leqslant \nu \, ;
  \\
  & \eta_{\mu\rho} \vect{\Omega}^{\rho}_{\nu}
  + \eta_{\nu\rho} \vect{\Omega}^{\rho}_{\mu}
  = 0
  \, . &
\end{alignat}

\section{Self-Dual 2-Forms}

Define some local 2-forms
\begin{alignat}{3}
  \tensor{S}^{1}_{\pm}
  & := \od\chi \wedge \od\zeta
  \pm i \cosh\chi \sin\zeta \cos\zeta \, \od\theta \wedge \od\varphi
  \, , \qquad &
  \tensor{C}^{1}_{\pm}
  &= \vect{\theta}^{0} \wedge \vect{\theta}^{1}
  \pm i \, \vect{\theta}^{2} \wedge \vect{\theta}^{3}
  && = l^{2} \cosh\chi \; \tensor{S}^{1}_{\pm}
  \, , \\
  \tensor{S}^{2}_{\pm}
  & := \od\chi \wedge \od\theta
  \mp i \cosh\chi \tan\zeta \, \od\zeta \wedge \od\varphi
  \, , \qquad &
  \tensor{C}^{2}_{\pm}
  &= \vect{\theta}^{0} \wedge \vect{\theta}^{2}
  \pm i \, \vect{\theta}^{3} \wedge \vect{\theta}^{1}
  &&= l^{2} \cosh\chi \cos\zeta \; \tensor{S}^{2}_{\pm}
  \, , \\
  \tensor{S}^{3}_{\pm}
  & := \od\chi \wedge \od\varphi
  \pm i \cosh\chi \cot\zeta \, \od\zeta \wedge \od\theta
  \, , \qquad &
  \tensor{C}^{3}_{\pm}
  &= \vect{\theta}^{0} \wedge \vect{\theta}^{3}
  \pm i \, \vect{\theta}^{1} \wedge \vect{\theta}^{2}
  &&= l^{2} \cosh\chi \sin\zeta \; \tensor{S}^{3}_{\pm}
  \, .
\end{alignat}
They satisfy
\begin{equation}
  * \tensor{S}^{j}_{\pm} = \pm i \, \tensor{S}^{j}_{\pm}
  \, , \qquad
  * \tensor{C}^{j}_{\pm} = \pm i \, \tensor{C}^{j}_{\pm}
  \, , \qquad
  (j = 1, 2, 3)
\end{equation}
and
\begin{alignat}{1}
  \Lied{\tensor{e}_{\alpha_{1}}} \tensor{S}^{1}_{\pm}
  &= \frac{i}{2} e^{i \varphi - i \theta} (\tensor{S}^{2}_{\pm} - S^{3}_{\pm})
  \, , \\
  \Lied{\tensor{e}_{\alpha_{1}}} \tensor{S}^{2}_{\pm}
  &= - \frac{i}{2} e^{i \varphi - i \theta}
    (
      \sec^{2} \zeta \; \tensor{S}^{1}_{\pm}
      - i \tan\zeta \; \tensor{S}^{2}_{\pm}
      + i \tan\zeta \; \tensor{S}^{3}_{\pm}
    )
  \, , \\
  \Lied{\tensor{e}_{\alpha_{1}}} \tensor{S}^{3}_{\pm}
  &= \frac{i}{2} e^{i \varphi - i \theta}
    (
      \csc^{2} \zeta \; \tensor{S}^{1}_{\pm}
      + i \cot\zeta \; \tensor{S}^{2}_{\pm}
      - i \cot\zeta \; \tensor{S}^{3}_{\pm}
    )
  \, , \\
  \Lied{\tensor{e}_{\alpha_{1} + 2 \alpha_{2}}} \tensor{S}^{1}_{\pm}
  &= - \frac{i}{2} e^{- i \theta - i \varphi}
    (\tensor{S}^{2}_{\pm} + \tensor{S}^{3}_{\pm})
  \, , \\
  \Lied{\tensor{e}_{\alpha_{1} + 2 \alpha_{2}}} \tensor{S}^{2}_{\pm}
  &= \frac{i}{2} e^{- i \theta - i \varphi}
    (
      \sec^{2} \zeta \; \tensor{S}^{1}_{\pm}
      - i \tan\zeta \; \tensor{S}^{2}_{\pm}
      - i \tan\zeta \; \tensor{S}^{3}_{\pm}
    )
  \, , \\
  \Lied{\tensor{e}_{\alpha_{1} + 2 \alpha_{2}}} \tensor{S}^{3}_{\pm}
  &= \frac{i}{2} e^{- i \theta - i \varphi}
    (
      \csc^{2} \zeta \; \tensor{S}^{1}_{\pm}
      + i \cot\zeta \; \tensor{S}^{2}_{\pm}
      + i \cot\zeta \; \tensor{S}^{3}_{\pm}
    )
  \, , \\
  \Lied{\tensor{e}_{\alpha_{2}}} \tensor{S}^{1}_{\pm}
  &= - e^{- i \varphi} \tanh\chi \sin\zeta \; \tensor{S}^{1}_{\pm}
  \mp e^{- i \varphi} \frac{\cos\zeta}{\cosh\chi} \, \tensor{S}^{2}_{\pm}
  - i e^{- i \varphi} \tanh\chi \cos\zeta \; \tensor{S}^{3}_{\pm}
  \, , \\
  \Lied{\tensor{e}_{\alpha_{2}}} \tensor{S}^{2}_{\pm}
  &= \pm \frac{\sec\zeta}{\cosh\chi} e^{- i \varphi} \tensor{S}^{1}_{\pm}
  \mp i \frac{\sin\zeta}{\cosh\chi} e^{- i \varphi} \tensor{S}^{3}_{\pm}
  \, , \\
  \Lied{\tensor{e}_{\alpha_{2}}} \tensor{S}^{3}_{\pm}
  &= i e^{- i \varphi} \tanh\chi \csc\zeta \cot\zeta \; \tensor{S}^{1}_{\pm}
  \pm i e^{- i \varphi} \frac{\cos\zeta \cot\zeta}{\cosh\chi}
    \tensor{S}^{2}_{\pm}
  - e^{- i \varphi} \tanh\chi \csc\zeta \; \tensor{S}^{3}_{\pm}
  \, , \\
  \Lied{\tensor{e}_{\alpha_{1} + \alpha_{2}}} \tensor{S}^{1}_{\pm}
  &= e^{- i \theta} \tanh\chi \cos\zeta \; \tensor{S}^{1}_{\pm}
  - i e^{- i \theta} \tanh\chi \sin\zeta \; \tensor{S}^{2}_{\pm}
  \mp \frac{\sin\zeta}{\cosh\chi} e^{- i \theta} \tensor{S}^{3}_{\pm}
  \, , \\
  \Lied{\tensor{e}_{\alpha_{1} + \alpha_{2}}} \tensor{S}^{2}_{\pm}
  &= i e^{- i \theta} \tanh\chi \sec\zeta \tan\zeta \; \tensor{S}^{1}_{\pm}
  + e^{- i \theta} \tanh\chi \sec\zeta \; \tensor{S}^{2}_{\pm}
  \mp i e^{- i \theta} \frac{\sin\zeta \tan\zeta}{\cosh\chi}
    \tensor{S}^{3}_{\pm}
  \, , \\
  \Lied{\tensor{e}_{\alpha_{1} + \alpha_{2}}} \tensor{S}^{3}_{\pm}
  &= \pm e^{- i \theta} \frac{\csc\zeta}{\cosh\chi} \; \tensor{S}^{1}_{\pm}
  \pm i e^{- i \theta} \frac{\cos\zeta}{\cosh\chi} \; \tensor{S}^{3}_{\pm}
  \, .
\end{alignat}
It follows that
\begin{alignat}{2}
  \Lied{\tensor{e}_{\alpha_{1}}} \tensor{C}^{1}_{\pm}
  &= \frac{i}{2} e^{i \varphi - i \theta} (
      \sec\zeta \; \tensor{C}^{2}_{\pm}
      - \csc\zeta \; \tensor{C}^{3}_{\pm}
    )
  \, , &
  \Lied{\tensor{e}_{\alpha_{1} + 2 \alpha_{2}}} \tensor{C}^{1}_{\pm}
  &= - \frac{i}{2} e^{- i \theta - i \varphi}
    (\sec\zeta \; \tensor{C}^{2}_{\pm} + \csc\zeta \; \tensor{C}^{3}_{\pm})
  \, , \\
  \Lied{\tensor{e}_{\alpha_{1}}} \tensor{C}^{2}_{\pm}
  &= - \frac{i}{2} e^{i \varphi - i \theta}
    (\sec\zeta \; \tensor{C}^{1}_{\pm} + i \, \tensor{C}^{3}_{\pm})
  \, , &
  \Lied{\tensor{e}_{\alpha_{1} + 2 \alpha_{2}}} \tensor{C}^{2}_{\pm}
  &= \frac{i}{2} e^{- i \theta - i \varphi}
    (\sec\zeta \; \tensor{C}^{1}_{\pm} - i \, \tensor{C}^{3}_{\pm})
  \, , \\
  \Lied{\tensor{e}_{\alpha_{1}}} \tensor{C}^{3}_{\pm}
  &= \frac{i}{2} e^{i \varphi - i \theta}
    (\csc\zeta \; \tensor{C}^{1}_{\pm} + i \, \tensor{C}^{2}_{\pm})
  \, , &
  \Lied{\tensor{e}_{\alpha_{1} + 2 \alpha_{2}}} \tensor{C}^{3}_{\pm}
  &= \frac{i}{2} e^{- i \theta - i \varphi}
    (\csc\zeta \; \tensor{C}^{1}_{\pm} + i \, \tensor{C}^{2}_{\pm})
  \, , \\
  \Lied{\tensor{e}_{\alpha_{2}}} \tensor{C}^{1}_{\pm}
  &= \mp \frac{e^{- i \varphi}}{\cosh\chi} \tensor{C}^{2}_{\pm}
  - i e^{- i \varphi} \tanh\chi \cot\zeta \; \tensor{C}^{3}_{\pm}
  \, , &
  \Lied{\tensor{e}_{\alpha_{1} + \alpha_{2}}} \tensor{C}^{1}_{\pm}
  &= -i e^{- i \theta} \tanh\chi \tan\zeta \; \tensor{C}^{2}_{\pm}
  \mp \frac{e^{- i \theta}}{\cosh\chi} \tensor{C}^{3}_{\pm}
  \, , \\
  \Lied{\tensor{e}_{\alpha_{2}}} \tensor{C}^{2}_{\pm}
  &= \pm \frac{e^{- i \varphi}}{\cosh\chi} \tensor{C}^{1}_{\pm}
  \mp i \frac{e^{- i \varphi} \cos\zeta}{\cosh\chi} \tensor{C}^{3}_{\pm}
  \, , &
  \Lied{\tensor{e}_{\alpha_{1} + \alpha_{2}}} \tensor{C}^{2}_{\pm}
  &= i e^{- i \theta} \tanh\chi \tan\zeta \; \tensor{C}^{1}_{\pm}
  \mp i \frac{e^{- i \theta} \sin\zeta}{\cosh\chi} \tensor{C}^{3}_{\pm}
  \, , \\
  \Lied{\tensor{e}_{\alpha_{2}}} \tensor{C}^{3}_{\pm}
  &= i e^{- i \varphi}
      \tanh\chi \cot\zeta \; \tensor{C}^{1}_{\pm}
      \pm i \frac{e^{- i \varphi} \cos\zeta}{\cosh\chi} \tensor{C}^{2}_{\pm}
  \, , \quad &
  \Lied{\tensor{e}_{\alpha_{1} + \alpha_{2}}} \tensor{C}^{3}_{\pm}
  &= \pm \frac{e^{- i \theta}}{\cosh\chi} \tensor{C}^{1}_{\pm}
  \pm i \frac{e^{- i \theta} \cos\zeta}{\cosh\chi} \tensor{C}^{3}_{\pm}
  \, .
\end{alignat}

\end{document}